\documentclass[prl,twocolumn,amsmath]{revtex4}
\usepackage{graphicx}
\begin{document}
\title{Multifractal current distribution in random diode networks}

\author{Haye Hinrichsen$^{1}$, Olaf Stenull$^{2,3}$, and Hans-Karl Janssen$^{2}$}
\affiliation{$^{1}$ Theoretische Physik, Fachbereich 8, Universit{\"a}t
	GH Wuppertal, 42097 Wuppertal, Germany}
\affiliation{$^{2}$ Institut f{\"u}r Theoretische Physik III, Heinrich-Heine-Universit{\"a}t, Universit{\"a}tsstr.~1, 40225 D{\"u}sseldorf, Germany}
\affiliation{$^{3}$ Department of Physics and Astronomy, David Rittenhouse Laboratory, 209 S.~33rd Street, Philadelphia, PA 19104-6396, USA}

\begin{abstract}
Recently it has been shown analytically that electric currents in a random diode network are distributed in a multifractal manner {[}O. Stenull and H. K. Janssen, Europhys.\ Lett.\ {\bf 55}, 691 (2001){]}. In the present work we investigate the multifractal properties of a random diode network at the critical point by numerical simulations. We analyze the currents running on a directed percolation cluster and confirm the field-theoretic predictions for the scaling behavior of moments of the current distribution. It is pointed out that a random diode network is a particularly good candidate for a possible experimental realization of directed percolation.
\end{abstract}

%\pacs{{\bf PACS numbers:} 05.70.Ln, 64.60.Ak, 64.60.Ht}]
% Explanation of PACS numbers:
% 05.70.Ln: Nonequilibrium and irreversible processes
% 64.60.Ak: order-disorder phase transitions / percolation transitions
% 64.60.Ht: dynamic critical phenomena                                                                         
\maketitle
\def\xvec{\text{\bf x}}

%
% INTRODUCTION TO DIRECTED PERCOLATION
%

\parskip 0mm 

Directed percolation (DP) is a simple model for directed connectivity in a random medium~\cite{MarroDickman,Hinrichsen00a}. It differs from ordinary isotropic percolation by the additional constraint that activity is restricted to percolate along a distinguished direction in space. Interpreting this direction as a temporal degree of freedom, directed percolation may also be viewed as a reaction-diffusion process $A \leftrightarrow 2A$, $A \to \emptyset$. In this process particle creation and removal compete one another, leading to a non-equilibrium phase transition form an active (percolating) into an absorbing (non-percolating) phase. Because of its robustness, DP plays the role of a standard universality class of nonequilibrium phase transitions which may be of a similar importance as the Ising model in equilibrium statistical mechanics.

\begin{figure}
\includegraphics[width=65mm]{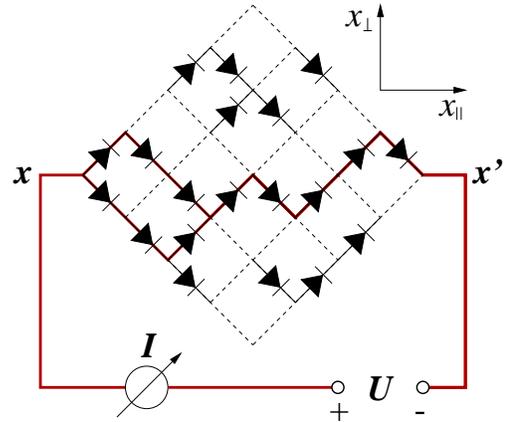}
\caption{
\label{FIGNETWORK}
(1+1)-dimensional directed bond percolation as a random-diode network on a tilted square lattice. The figure shows a realization of randomly distributed diodes with a conducting path from $x$ to $x'$.
} \vspace{-2mm}
\end{figure}

One of the simplest realizations of DP is directed bond percolation on a (tilted) hypercubic lattice. In this model neighboring sites are connected by bonds which are conducting with probability $p$ and impermeable otherwise. In contrast to isotropic percolation, the bonds conduct only in one direction and may be realized by randomly distributed diodes, as shown in Fig.~\ref{FIGNETWORK}. Obviously, the effective conductivity of the system along the distinguished direction depends on the percolation probability $p$. If $p$ is small, all cluster of connected sites are finite so that the conductivity vanishes in the limit of large distances. On the other hand, if $p$ exceeds a certain critical threshold $p_c$, there is a finite probability to find a directed path between two points, leading to a finite resistance when averaged over many independent samples. At the critical threshold $p=p_c$, the system undergoes a continuous phase transition where clusters of conducting paths display a fractal structure. In this case the currents running on the cluster are distributed in a nontrivial way.

% SCALING PROPERTIES OF DP

The DP transition can be described by field-theoretic renormalization group methods~\cite{ReggeonFT,Janssen81}. In this framework the critical behavior near the transition can be characterized by simple scaling laws which involve three independent critical exponents. The order parameter in the percolating phase is the density of sites belonging to an infinite cluster or -- using the reaction-diffusion language --  the density of active sites. Approaching the transition, this density scales as $\rho \sim (p-p_c)^\beta$, where $\beta$ is the critical exponent associated with the order parameter. Similarly, the correlation length along the distinguished direction scales as $\xi_\parallel\sim |p-p_c|^{-\nu_\perp}$, while the correlation length perpendicular to this direction diverges as $\xi_\perp\sim |p-p_c|^{-\nu_\perp}$. The exponents $\nu_\parallel$ and $\nu_\perp$ are generally different since introducing a distinguished direction in space breaks the symmetry between parallel and perpendicular degrees of freedom. In a two-loop approximation, these exponents are given by~\cite{Janssen81,ExponentsFT}
\begin{equation}
\label{pile}
\begin{split}
\beta &= 1-\epsilon/6-0.01128\,\epsilon^2 + O(\epsilon^3)\\
\nu_\parallel &= 1+\epsilon/12+0.02238\,\epsilon^2  + O(\epsilon^3)\\
\nu_\perp &= 1/2+\epsilon/16+0.02110\,\epsilon^2 + O(\epsilon^3)
\end{split}
\end{equation}
where $\epsilon=4-d$. The exponents have also been estimated by various numerical methods (see Table~\ref{TableStandardExponents}).
The triplet of exponents $(\beta,\nu_\parallel, \nu_\perp)$ labels the universality class of $d$+1 dimensional DP, while other critical exponents are usually related to these standard exponents by simple scaling relations. This type of scaling invariance, where only a finite number of independent exponents is involved, is referred to as {\em simple scaling}.

% TWOPOINT CURRENTS AND MULTIFRACTALITY

Let us now consider an electric current running on a directed percolation cluster according to Kirchhoff's laws. By introducing such a current the theory is extended by an additional physical concept with non-trivial properties. In fact, even though the cluster itself is characterized by simple scaling laws, the currents {\em on} the cluster turn out to be distributed in {\em multifractal} manner, meaning that all moments scale with individual exponents which cannot be related to the standard exponents by simple scaling laws. In the case of isotropic percolation, this phenomenon was first conjectured by Rammal {\sl et al.}~\cite{rammal_etal_85} as well as by Arcangelis {\sl et al.}~\cite{arcangelis_etal_85}. In the meantime these results have been substantiated by field-theoretic renormalization group studies~\cite{park_harris_lubensky_87,ResistorNetwork} and numerical simulations~\cite{rammal_etal_85,Batrouni,Barthelemy}. Recently, Stenull and Janssen~\cite{StenullJanssen2} investigated the directed case. By employing field theoretic methods they determined the critical exponents of the moments of the current distribution. Their results clearly indicate multifractality. The purpose of the present work is to verify these predictions quantitatively by extensive numerical simulations. To this end we analyze a random diode network as shown in Fig.~\ref{FIGNETWORK}, which is a special case of the model considered in~\cite{StenullJanssen2}. Again we would like to stress that multifractality is not observed as long as one is concerned exclusively with the underlying DP process. An additional process, here the transport of electrical currents, is necessary for the emergence of multifractality.

% DEFINITION OF MOMENTS

The moments of the current distribution in a $d$+1-dimensional DP process are defined as follows. Let us denote by $x_\parallel$ the coordinate along the distinguished direction (=time) and by $x_\perp$ the vector of $d$ perpendicular coordinates (=space). An important quantity in the theory of DP is the pair-connectedness function $C(x_\parallel,x_\perp; x'_\parallel,x'_\perp)$, which is the probability to find a conducting path running from the point $(x_\parallel,x_\perp)$ to the point $(x'_\parallel,x'_\perp)$. The union of all possible paths in a given realization of conducting and non-conducting bonds is called the {\em backbone} of the correlation function. Note that the backbone is invariant under reversal of all diodes. In the field-theoretic framework this invariance is known as the so-called rapidity reversal symmetry~\cite{RapidityReversal}.

\begin{table}
\begin{ruledtabular}
\begin{tabular}{c|ccc}
exponent & $d=3$ & $d=2$ & $d=1$\\
\hline
$\beta$ & $0.81(1)$ & $0.583(4)$ & $0.276\,486(8)$ \\
$\nu_\parallel$ & $1.105(5)$ & $1.295(6)$ & $1.733\,847(6)$ \\
$\nu_\perp$ & $0.581(5)$ & $0.733(4)$ & $1.096\,584(4)$ \\
\end{tabular}
\end{ruledtabular}
\caption{\label{TableStandardExponents}
Estimates for the critical exponents of DP.
} \vspace{-4mm}
\end{table}

\begin{figure}
\includegraphics[width=85mm]{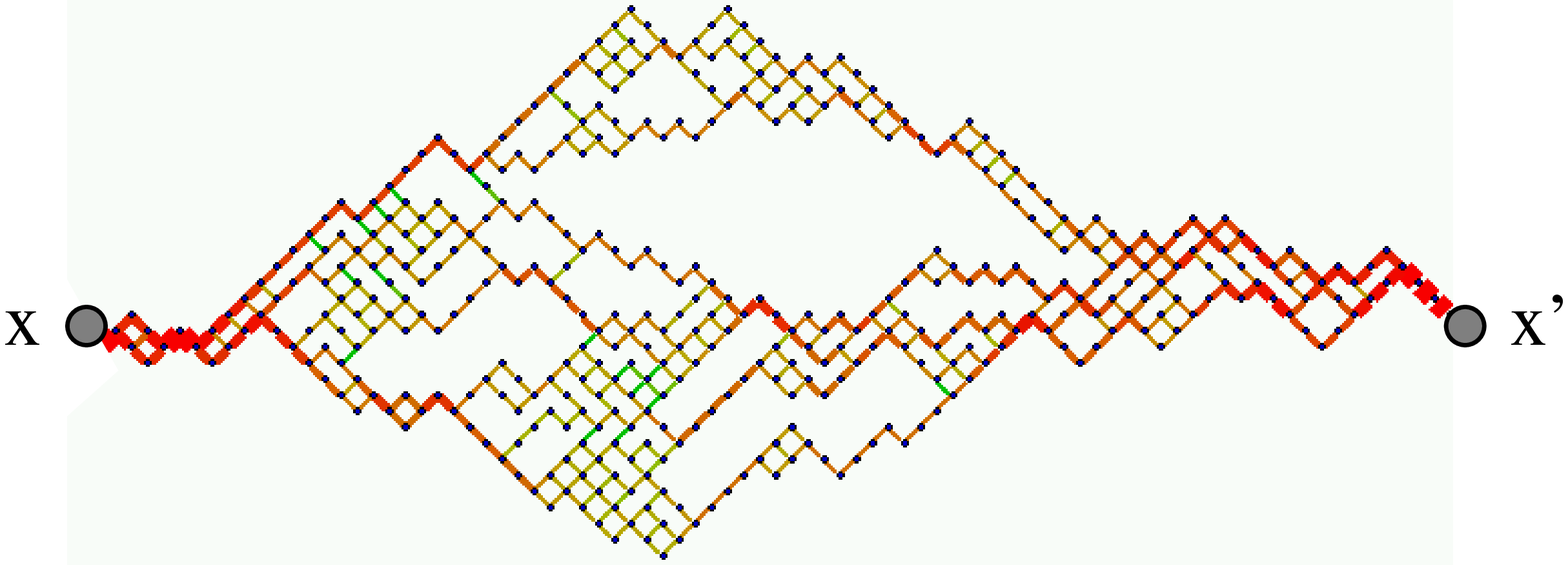}
\caption{
\label{FIGDEMO}
Example of the current distribution on a backbone running from $\xvec$ to $\xvec'$. The thickness of the bonds represents the intensity of the current.
} \vspace{-2mm}
\end{figure}

We now assume the diodes to be ideal in the sense that they are perfectly isolating if the applied voltage is negative while they have a fixed resistance $R$ if the voltage is positive. If we apply an external voltage at the two points $\xvec=(x_\parallel,x_\perp)$ and $\xvec'=(x'_\parallel,x_\perp)$ there will be a current running on the backbone which can be computed by solving Kirchhoff's laws. Obviously the currents are distributed in a non-trivial way, especially when the backbone consists of many nested loops, as shown in Fig.~\ref{FIGDEMO}. This current distribution has multifractal properties which can be studied by considering the moments
\begin{equation}
M_\ell = \sum_b (I_b/I)^{2 \ell} \,,
\end{equation}
where the sum runs over all bonds $b$ of the backbone while $I_b$ and $I$ denote the current passing bond $b$ and the total current, respectively. Each of these moments has a specific physical meaning. For example, $M_0$ is just the total number of bonds of the backbone while $M_1$ is the total resistance between the two points in units of $R$. $M_2$ is the second cumulant of the resistance fluctuations and may be considered as a measure of the noise in a given realization. Finally, $M_\infty$ is the number of so-called red bonds which carry the full current $I$. These bonds play a special role since they are shared by all possible paths.

%
% RECENT FIELD-THEORETIC RESULTS

At criticality the moments obey a power-law~\cite{StenullJanssen2}
\begin{equation}
M_\ell \sim (x'_\parallel -x_\parallel)^{\psi_\ell/\nu_\parallel} \,.
\end{equation}
Multifractal behavior manifests itself in a nonlinear dependence of the exponents $\psi_\ell$ on their index $\ell$. The exponents $\psi_\ell$, just as the geometrical exponents compiled in Eq.~(\ref{pile}), are known in a two-loop approximation. Thus, at least in principle, we can gather the $\epsilon$ expansion for the exponents $\psi_\ell/\nu_\parallel$ from the literature. This $\epsilon$ expansion, however, would hardly be useful for comparison to our numerical data because it is valid only asymptotically in the vicinity of $d_c =4$. To produce reliable predictions for $d=3$ or even $d=2$ the $\epsilon$ expansion has to be improved by incorporating additional information. Here, we carry out a so-called rational approximation. To be specific, we incorporate the rigorous feature $\psi_\ell/\nu_\parallel (d=0) =1$ by supplementing the $\epsilon$ expansion with an appropriate third order term. The so obtained result reads:
\begin{equation}
\label{monsterExponent}
\frac{\psi_\ell}{\nu_\parallel} = 1 - \epsilon \left( 4 - \epsilon \right) \left\{ a_\ell + \epsilon \left[ b_\ell + c_\ell \, \ln \left( \frac{4}{3} \right) \right] \right\} 
+ O \left( \epsilon^4 \right)
\end{equation} 
where the $a_\ell$, $b_\ell$ and $c_\ell$ are $\ell$-dependent coefficients taking the values listed in Table~\ref{tab:coeffs}.
\begin{figure}
\includegraphics[width=85mm]{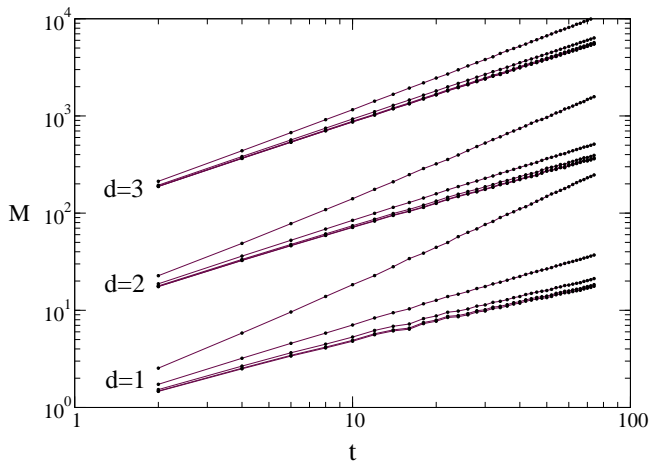}
\caption{
\label{FIGDATA}
Simulation results for the moments $M_\ell$ as a function of time in 1+1,2+1, and 3+1 dimensions. The data points for 2+1 and 3+1 dimensions are shifted vertically by a factor of 10 and 100, respectively. Each group of lines shows the moments $M_0,\ldots,M_5$ from top to bottom. The data points for $M_4$ and $M_5$ are almost identical.
} \vspace{-2mm}
\end{figure}

In order to estimate the exponents numerically, we determine the current distribution averaging over an ensemble of 5000 independent realizations of a conducting backbone in 3+1, 2+1, and 1+1 dimensions. In principle the current distribution can be computed by solving Kirchhoff's laws at all nodes of the backbone. In the case of isotropic percolation, where the network consists only of resistors, one obtains a set of linear equations which can be solved exactly. In the present case of a diode-resistor network, however, the resulting equations are {\em nonlinear} making it impossible to compute the currents analytically. Thus, in order to compute the currents, we have to apply an iterative approximation procedure. To this end each site of the backbone network is connected to a grounded capacitor $C$. For a given set of voltages at the lattice sites, we can compute the actual difference between incoming and outgoing currents. This difference then changes the voltage of the capacitor at site $i$ by
\begin{equation}
\label{Iteration}
\frac{dU_i}{dt} = \frac{1}{C} \Bigl( I_{\text{incoming}} - 
I_{\text{outgoing}}  \Bigr) \,.
\end{equation}

After sufficiently long time this nonlinear dynamical system reaches a stationary state where incoming and outgoing currents balance each other. Therefore, by iterating Eq.~(\ref{Iteration}), the stationary current distribution can be approximated accurately, mainly limited by the available CPU time and the precision of floating-point arithmetics. Moreover, the convergence during the iteration procedure can be controlled easily by monitoring the quantity $\sum_b|\Delta I_b|$, which should tend to zero as the stationary state is approached. Using Newton's iteration scheme, this quantity actually decreases until it reaches a value of about $10^{-10}$, which is essentially determined by the limited machine precision. At this point we stop the iteration procedure and compute the moments $M_\ell$, assuming that the errors caused by the iteration procedure are much smaller than the statistical and finite-size errors generated by the DP process itself.

In the simulations we observe that the average size of the backbone increases rapidly with $x'_\parallel-x_\parallel$, especially in 3+1 dimensions. While the numerical effort for generating conducting backbones between two given points plays a minor role, the algorithm is mainly limited by the CPU time needed for the dynamical approximation and by the memory capacity for storing the configuration of the backbone, the currents, and the voltages. Because of these limitations, we could only go up to distances of $x'_\parallel-x_\parallel \leq 75$ in 3+1, 2+1 and 1+1 dimensions. Clearly, in this limited range corrections to scaling still play an important role. Nevertheless we find that the moments exhibit very clean power laws (see Fig.~\ref{FIGDATA}), which allows us to estimate the critical exponents $\psi_\ell$ in the limit $x'_\parallel-x_\parallel \to \infty$ by standard extrapolation techniques. These estimates are summarized in Table~\ref{TableExponents}.

\begin{table}
\begin{ruledtabular}
\begin{tabular}{c|ccccc}
$\quad \ell \quad $ & $0$ & $1$ & $2$ & $3$ & $4$ \\ 
\hline
$ a_\ell $ & $-\frac{1}{48}$ & $\frac{1}{96}$ & $\frac{7}{384}$ & $\frac{31}{1536}$ & $\frac{127}{6144}$ \\
$ b_\ell $ & $-\frac{109}{13824}$ & $\frac{115}{27648}$ & $\frac{149053}{19906560}$ & $\frac{11739503}{1337720832}$ & $\frac{23399081351}{2476694568960}$\\
$ c_\ell $ & $-\frac{161}{6912}$ & $\frac{47}{13824}$ & $\frac{211}{110592}$ & $-\frac{491}{1769472}$ & $-\frac{17377}{9437184}$\\
\end{tabular}
\end{ruledtabular}
\caption{\label{tab:coeffs}
The coefficients $a_\ell$, $b_\ell$ and $c_\ell$ appearing in Eq.~(\ref{monsterExponent}).} \vspace{-2mm}
\end{table}

\begin{table}
\begin{ruledtabular}
\begin{tabular}{c|ccc}
exponent & $d=3$ & $d=2$ & $d=1$\\
\hline
$\psi_0/\nu_\parallel$ & 1.10(2) & 1.24(3) & 1.32(3) \\
$\psi_1/\nu_\parallel$ & 0.95(1) & 0.88(2) & 0.82(2)  \\
$\psi_2/\nu_\parallel$ & 0.92(1) & 0.81(3) & 0.67(3) \\
$\psi_3/\nu_\parallel$ & 0.91(1) & 0.79(3) & 0.63(3) \\
$\psi_4/\nu_\parallel$ & 0.91(1) & 0.78(3) & 0.62(3) 
\end{tabular}
\end{ruledtabular}
\caption{\label{TableExponents}
Numerical estimates of the critical exponents.
} \vspace{-2mm}
\end{table}

\begin{figure}
\includegraphics[width=66mm]{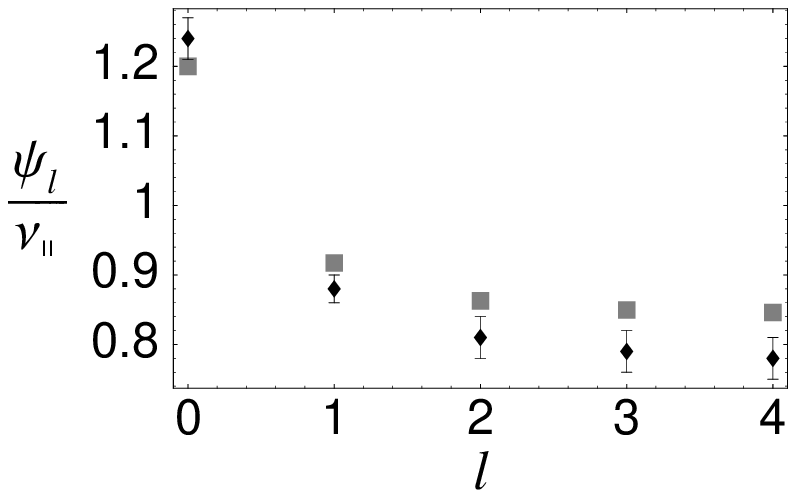}
\includegraphics[width=70mm]{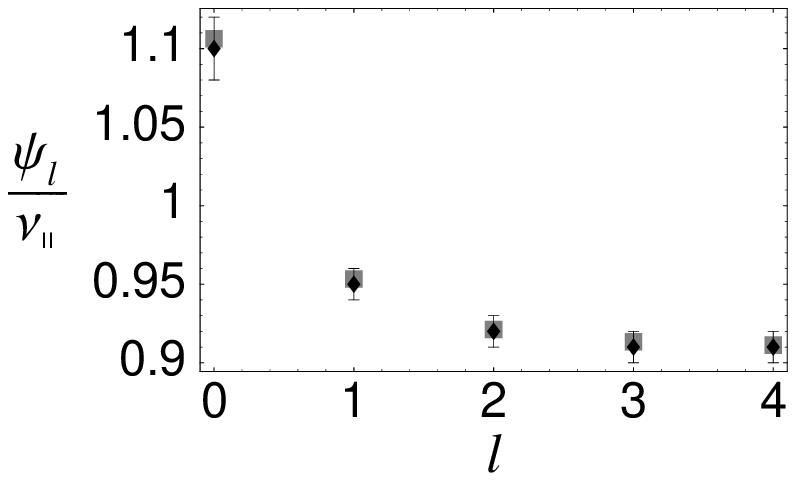}
\vspace{-5mm}
\caption{
\label{FIGRESULTS}
Comparison of simulation and field theory for $d=2$ (top) and $d=3$ (bottom). The field theoretic results are indicated by squares. The numerical ones are symbolized by diamonds with error bars.} \vspace{-2mm}
\end{figure}

Figure~\ref{FIGRESULTS} shows the field-theoretic estimates as given in Eq.~(\ref{monsterExponent}) compared to the numerical results. The error bars include the statistical error of the simulation combined with the expected error of the extrapolation procedure. In 3+1 dimensions we find excellent
agreement. The numerical values and the field theoretic predictions almost coincide. As expected, the agreement is less pronounced in 2+1 dimensions. The analytic estimates lie slightly outside the error bars of the simulations. Though not quantitatively correct, field theory yet predicts the right shape of the dependence of $\psi_\ell / \nu_\parallel$ on $\ell$. In 1+1 dimension, the field theory fails to predict the numerical values of the multifractal exponents accurately. We therefore refrain from showing the respective graph. All in all, our simulations support the validity of the field theoretic results, at least for dimensions to that the $\epsilon$ expansion can be continued reliably. Our simulations clearly confirm the existence of multifractality in current distributions running on DP clusters.

As in the field-theoretic case, we can check the consistency of our results. For example, as shown in~\cite{StenullJanssen2}, the zeroth moment in a ($d$+1)-dimensional system, which is just the mass of the backbone, can be expressed in terms of the backbone dimension $D_B=d+1-2\beta/\nu_\perp$ (cf.\ Ref.~\cite{Backbonedim}) by
\begin{equation}
\psi_0 = \nu_\perp (D_B -1 + z)\,,
\end{equation}
where $z=\nu_\parallel/\nu_\perp$. Using the values listed in Table~\ref{TableStandardExponents}, we obtain for $\psi_0/\nu_\parallel$ the values 1.11 in three, 1.23 in two, and 1.307 in one dimension, respectively. These values are in perfect agreement with the estimates listed in Table~\ref{TableExponents}.

To summarize, we have studied the moments of currents running in a random diode network at the critical point by numerical simulations. Our results fully confirm recent field-theoretic results, supporting the claim that currents running on a critical DP cluster indeed show multifractal behavior.

We would like to point out that a random diode network might be a good candidate for an experimental realization of directed percolation. In particular the total resistance $M_1$ is easy to measure. In an experimental realization various idealized assumptions of the model shown in Fig.~\ref{FIGNETWORK} can be relaxed. For example, not all diodes need to be polarized along the distinguished direction, rather it is sufficient for them to be {\it preferentially oriented} in a statistical sense. In addition, the resistance of conducting diodes may be stochastically distributed. Nevertheless the critical exponents of such a system at the transition are expected to be the same as in DP. Even inhomogeneities, which cannot be avoided in any experiment, are not expected to play a role since they would result into an {\it annealed} noise which is irrelevant under renormalization group transformations (see Ref.~\cite{Hinrichsen00b}). Because of these favorable properties, random diode networks are particularly suitable for a possible experimental realization of directed percolation. It should be emphasized that so far {\it no} experiment is known where the critical exponents of directed percolation could be verified in a reliable way~\cite{Hinrichsen00b,Hinrichsen99}. Therefore, the experimental realization of a random diode network would be a particularly challenging task.

H. K. J. acknowledges the support by the Sonderforschungsbereich 237 ``Unordnung und gro{\ss}e Fluktuationen'' of the Deutsche Forschungsgemeinschaft. O. S. is grateful for the support by the Emmy-Noether-Programm of the Deutsche Forschungsgemeinschaft.

\vspace{-4mm}

%====================================================================

\end{document}